\documentclass[nofootinbib,prd,twocolumn,showpacs,showkeys,preprintnumbers]{revtex4-1}
\usepackage{hyperref,amssymb,amsmath,mathrsfs,bm,graphicx}
\begin{document}
\title {The double polytrope for anisotropic matter: Newtonian Case}
\author{G. Abell\'an}
\email{gabriel.abellan@ciens.ucv.ve}
\affiliation{Escuela de F\'\i sica, Facultad de Ciencias, Universidad Central de Venezuela, Caracas 1050, Venezuela}
\author{E. Fuenmayor}
\email{ernesto.fuenmayor@ciens.ucv.ve}
\affiliation{Escuela de F\'\i sica, Facultad de Ciencias, Universidad Central de Venezuela, Caracas 1050, Venezuela}
\author{L. Herrera}
\email{lherrera@usal.es}
\affiliation{Instituto Universitario de F\'isica
Fundamental y Matem\'aticas, Universidad de Salamanca, Salamanca 37007, Spain}

\begin{abstract}
A general formalism developed  few years ago  to model polytropic Newtonian stars with anisotropic pressure is applied to model stars for which, both, radial and tangential pressure satisfy polytropic equations of state. We obtain the corresponding Lane--Emden equation, and integrate it for a  wide range of values of the parameters, thereby illustrating the effects of the pressure anisotropy on stellar objects. We calculate Chandrasekhar mass for a white dwarf and compare it with previous results. Finally, prospective applications  to some astrophysical scenarios are discussed.
\end{abstract}
\date{\today}
\maketitle

\section{Introduction}\label{Intro}
In the theory of stellar structure, in the context of Newtonian gravity, the polytropic equation of state has played a very important role \cite{Abramowicz1983,Chandrasekhar1939,Horedt2004,Kippenhahn2012}. The reasons behind the success  of this approach lie, on the one hand,  in  the simplicity of the equation of state and the ensuing structure equation (the Lane-Emden equation), and on the other hand, on the fact that the polytropic equation of state may be applied to a wide range of very  different astrophysical scenarios. Although we are restricting in this work to the Newtonian case, it should be mentioned that for extremely compact objects, a relativistic theory of gravitation has to be used (e.g.  general relativity) which requires a suitable adapted  approach, described in many references (see for example \cite{Bludman1973,Herrera:2003ua,Lai:2008cw,Nilsson:2000zg,Tooper1964,Tooper1965,Tooper1966,Herrera:2013fja,Thirukkanesh2012,Maeda:2002br,Kim2017,Mardan2020,Bhatti2019,Bhatti2020,Roy2019,Maharaj2015,Azam2017,Moussa2017,Harko2016,Herrera2016,Herrera2014} and references therein).

It is worth noticing that in most applications it is usual to assume that  the fluid distribution  satisfies the Pascal principle (principal stresses equal),  i.e. the pressure is assumed to be isotropic. However, we know nowaday that the isotropic pressure condition may be too stringent, and furthermore  the presence of local anisotropy is caused by a large variety of physical phenomena expected to be present in compact objects \cite{Herrera:1997plx}.

For  the isotropic pressure case, the polytropic equation of state reads
\begin{equation}\label{intro01}
    P \;=\; K\rho^\gamma \;=\; K\rho^{1+1/n}\;,
\end{equation}
where $P$ and $\rho$ denote the isotropic pressure and the density respectively, and  $K$,  $\gamma$,  and $n$  are  the polytropic constant, the polytropic exponent, and the polytropic index, respectively. To any duplet of these two parameters ($K, \gamma$ or $K, n$) corresponds a specific model of the star. When polytropic constant $K$ is fixed and can be calculated from natural constants, the polytropic equation of state may be  used to model a completely degenerate Fermi gas in the nonrelativistic ($n=5/3$) and relativistic limit ($n=4/3$). In this case, polytropes provide a tool of modelling compact objects as white  dwarfs and allow to obtain in a rather direct way the Chandrasekhar mass limit. On the other hand, when $K$ is a free parameter, constant for a specific star, polytropes can be used to model an isothermal ideal gas. Such models are   relevant in connection with the Schonberg-Chandrasekhar limit \cite{Kippenhahn2012}.

The theory of Newtonian polytropes for anisotropic matter was fully developed in \cite{Herrera:2013dfa} (for the relativistic version see \cite{Herrera:2013fja}), here we shall follow closely that approach, complementing it by assuming that both pressures (radial and tangential) satisfy a polytropic equation of state. Doing so the ensuing Lane--Emden equation may be integrated and the models would depend on the specific values of the parameters.

The manuscript is organized as follows. In the next section we outline the general method for treating Newtonian polytropes for anisotropic matter described in \cite{Herrera:2013dfa}. In order to integrate de ensuing Lane--Emden equation we need an additional ansatz, which in our case consists in assuming that  not only the radial pressure, but also the tangential one, obeys a polytropic equation of state. This is done in section III. In section IV, the corresponding Lane--Emden equation is integrated numerically for a specific range of values of the parameters. Finally, a discussion of the results is presented in the last section.

\section{Anisotropic Polytropes}\label{sec:2}
We shall here expose the basic ideas to treat the polytrope for anisotropic matter described in \cite{Herrera:2013dfa}, although we shall express the ensuing equations in a slightly different form.

For an anisotropic fluid  the hydrostatic equilibrium equation in spherical coordinates is given by
\begin{equation}\label{sec2_01}
    \frac{dP_r}{dr} \;=\; -\frac{d\phi}{dr}\rho + \frac{2}{r}\Delta\;,
\end{equation}
where $\phi$ is the gravitational potential and $\Delta = P_\perp - P_r$ the anisotropy factor. This equation can be considered as the Newtonian limit of Tolman--Opphenheimer--Volkoff equation for anisotropic matter. It is important to mention that due to the spherical symmetry, only two principal stresses may be unequal, i.e.  $P_\theta = P_\varphi = P_\perp$ and $P_r \neq P_\perp$.

On the other hand,  the Poisson equation reads
\begin{equation}\label{sec2_02}
    \frac{1}{r^2} \frac{d}{dr}\left( r^2 \frac{d\phi}{dr} \right)
    \; =\; 4\pi G\rho\;,
\end{equation}
where  $G$ denotes the gravitational constant. Considering that the fluid satisfies  a polytropic equation of state for the radial pressure $P_r$, then combining the Eqs. (\ref{intro01}) and (\ref{sec2_01}), we obtain
\begin{equation}\label{sec2_03}
    \gamma K \rho^{\gamma - 2} \frac{d\rho}{dr} \;=\; -\frac{d\phi}{dr} + \frac{2}{r\rho}\Delta\;. 
\end{equation}
Two different cases can be considered from now on: $\gamma\neq1$ and $\gamma=1$.

Let us first consider  $\gamma\neq 1$ and integrate from $r=r_c=0$ up to some arbitrary $r$, one obtains
\begin{equation}\label{sec2_04}
    \phi - \phi_c \;=\; -K(1+n)(\rho^{1/n} - \rho^{1/n}_c) + \int^r_0 \frac{2\Delta}{x\rho}dx\;,
\end{equation}
where we have used the relation between the exponent and the index of the polytrope $\gamma = 1 + 1/n$. The subscript $c$ denotes the value of the respective physical quantity evaluated at $r=0$. Next, we substitute Eq. (\ref{sec2_04}) in the Poisson equation (\ref{sec2_02}) to obtain
\begin{equation}\label{sec2_05}
    K(1+n)\,\nabla^2 \rho^{1/n} - \frac{1}{r^2}\frac{d}{dr}\! \left( r^2 \frac{2\Delta}{r\rho} \right)\!
    \;=\; -4\pi G\rho\,.
\end{equation}
In order to solve this equation it is convenient to rewrite it in dimensionless form, for this purpose the following quantities are defined
\begin{eqnarray}\label{sec2_06}
    z &=& Ar\;,\nonumber\\
    A^2 &=& \varepsilon\, \frac{4\pi G \rho_c^{1-1/n}}{K(1+n)}\;,\\
    \rho &=& \rho_c\, \omega^n\;.\nonumber
\end{eqnarray}
The parameter $\varepsilon$ can take two values depending on polytropic index range. If $n>-1$, $\varepsilon = +1$; if $n<-1$,  $\varepsilon = -1$. With these conventions the polytropic equation of state reads
\begin{equation}\label{sec2_07}
    P_r \;=\; K\rho^{1+1/n} \,=\, K\rho_c^{1+1/n}\omega^{1+n} \,=\, P_c\,\omega^{1+n}\;.
\end{equation}
Using Eq. (\ref{sec2_07}) and the conventions (\ref{sec2_06}), the Eq. (\ref{sec2_05}) takes the form
\begin{eqnarray}\label{sec2_08}
    \omega''+ \frac{2}{z}\omega' - \frac{2}{P_c(1+n)z\omega^n}\!
    \left(\! \Delta' + \frac{\Delta}{z} - n\frac{\omega'}{\omega}\Delta \! \right)\! = -\varepsilon\omega^n .\;\; \nonumber\\
\end{eqnarray}
Here, primes denote differentiation with respect to the adimensional coordinate $z$. Note that the variable $\omega$ is also dimensionless, in particular at the center of the system we have $\omega(0) = 1$. Also, note that in the isotropic limit $\Delta\to 0$ we recover the well known Lane--Emden equation \cite{Kippenhahn2012,Herrera:2013dfa}.

The case $n=-1$ corresponds to an equation of state of the form $P_r = K$, describing a constant pressure system (if $\rho \neq 0$). This of course  is not consistent with the boundary condition implying that the radial pressure vanishes on the boundary of the fluid distribution, accordingly  this case is ruled out.

Let us now consider the case  $n=\pm\infty$ ($\gamma=1$), producing  $P_r = K\rho$. Then  substituting this equation in (\ref{sec2_01}), integrating between $r=0$ and an arbitrary $r$, and using the Poisson equation (\ref{sec2_02}) we find
\begin{equation}\label{sec2_09}
    \omega'' + \frac{2}{z}\omega' + \frac{2\,e^\omega}{P_c z} \left( \Delta' + \frac{\Delta}{z}
    + \Delta\,\omega' \right) \;=\; e^{-\omega}\;,
\end{equation}
where we have used the following redefinitions in order to adimensionalize the equation
\begin{eqnarray}\label{sec2_10}
    z &=& \alpha\, r\;,\nonumber\\
    \alpha^2 &=& \frac{4\pi G \rho_c}{K}\;,\\
    \rho &=& \rho_c\, e^{-\omega}\;.\nonumber
\end{eqnarray}
The equation (\ref{sec2_09}) reduces to the respective isotropic Lane--Emden equation when $\Delta\to 0$, as it should be. Please note that in Eq. (\ref{sec2_08}) we have $P_c = K\rho_c^{1+1/n}$ but in Eq. (\ref{sec2_09}) we define $P_c = K\rho_c$.

Both Eqs. (\ref{sec2_08}) and (\ref{sec2_09}) are second order differential equations, so we must give two conditions in order to solve. Note that for $n\neq -1,\,\pm\infty$ we have $\omega(0) = 1$ and for $n=\pm\infty$, we use $\omega(0) = 0$. Now, in order to find the condition over the first derivative of $\omega$, we  integrate the Poisson equation (\ref{sec2_02})
\begin{equation}\label{sec2_11}
    \frac{d\phi}{dr} \;=\; \frac{4\pi G}{r^2}\! \int^r_0\! x^2\rho\; dx\;.
\end{equation}
For $n \neq -1,\,\pm\infty$ we use Eqs. (\ref{sec2_06}) and after substitution  in (\ref{sec2_03}) and integration, we find
\begin{equation}\label{sec2_12}
    \omega' \;=\; -\frac{\varepsilon}{z^2}\int^z_0 \omega^n x^2 dx \;+\; \frac{2}{P_c(1+n)}
    \left(\! \frac{\Delta}{z\omega^n}\! \right) .
\end{equation}
On the other hand, for $n = \pm\infty$, and by doing similar steps we obtain
\begin{equation}\label{sec2_13}
    \omega' \;=\; \frac{1}{z^2}\int^z_0 e^{-\omega} x^2 dx \;-\; \frac{2\Delta}{P_c\, z}e^{\omega}\,.
\end{equation}
Therefore, in the limit $z\to 0$ we find that
\begin{eqnarray}
    \omega'(0) &=&  \frac{2}{P_c(1+n)}\;\lim_{z\to 0}\;
    \frac{\Delta}{z}\;, \hspace{.3cm}  \mbox{for} 
    \hspace{.2cm} n \neq -1,\,\pm\infty\,,\;\;\;\;\;\label{sec2_14} \\
    \omega'(0) &=& -\frac{2}{P_c} \;\lim_{z\to 0}\; \frac{\Delta}{z}  
    \;, \hspace{1.1cm} \mbox{for} \hspace{.2cm} n = \pm\infty\,.\;\;\;\label{sec2_15}
\end{eqnarray}
Note that the expressions for the first derivatives impose a regularity condition over the anisotropy. 

As is evident from either  Eq. (\ref{sec2_08}) or Eq. (\ref{sec2_09}), in order to integrate them and obtain specific models, additional information has to be provided (we have one equation for two unknown variables).  As we mentioned before, we shall close the system by  considering  that the tangential pressure  (as well as the radial one)  satisfies  a polytropic equation of state. This ansatz is completely equivalent to giving the anisotropy factor. The structure equation as well as the rationale behind such an assumption will be  deployed and explained in the next section.

\section{The double polytrope}\label{sec:3}
Thus  we will assume that, both, radial and tangential pressures obey  a polytropic equation of state. Besides the need to provide additional information in order to obtain specific models, there is a simple idea behind the above mentioned assumption. Indeed, for a small anisotropy factor, if we assume that radial pressure satisfies a polytropic equation of state, then it is physically meaningful to impose a similar equation of state  for the tangential pressure. Therefore our assumption allows us to continuously connect our models through the variation of the parameters, including the isotropic pressure case, in a very simple way, with the certainty that, at least,  for non very large anisotropies the models should behave physically well.

In order to make the discussion easier, we shall consider separately different ranges of values of the parameters.

\subsection{Case 1: Both polytropes with $n\neq -1,\pm\infty$}
We start with our main  assumption, i.e.  both pressures obey a polytropic equation of  the form (\ref{intro01}), so that
\begin{eqnarray}
    P_r\, &=& \, K_r\rho^{\gamma_r} \;=\; P_{rc}\, \omega^{1+n_r}\,, \label{sec03_01}\\
    P_\perp\, &=& \, K_\perp \rho^{\gamma_\perp} \;=\; P_{\perp c}\, \omega^{1+n_\perp} \,. \label{sec03_02} 
\end{eqnarray}
 In the former expressions we used the change of variables $\rho = \rho_c\omega^n$ in (\ref{sec2_06}) but considering that there are two  polytropic indexes ($n_r, n_\perp$), one  for each polytrope . Clearly, from Eqs. (\ref{sec03_01}) and (\ref{sec03_02}) we have $P_{rc} = K_r\rho_c^{1+1/n_r}$ and $P_{\perp c} = K_r\rho_c^{1+1/n_\perp}$.

Using  (\ref{sec03_01}) and (\ref{sec03_02}) we write the anisotropy factor resulting from this
\begin{equation}\label{sec03_03}
    \Delta = P_\perp - P_r = P_{\perp c}\, \omega^{1+n_\perp} -
    P_{rc}\, \omega^{1+n_r}.
\end{equation}
Then, the regular condition at  $r=0$, implying that the anisotropy factor must be zero in the center $\Delta(0) = 0$, produces
\begin{equation}\label{sec03_04}
    P_{\perp c} \;=\; P_{rc} \hspace{.5cm} \longleftrightarrow \hspace{.5cm} \frac{K_\perp}{K_r} 
        \;=\; \frac{\rho_c^{1/n_r}}{\rho_c^{1/n_\perp}}\;.\;\;
\end{equation}
From this expression we see that choosing $K_\perp = K_r$ necessarily implies $n_\perp = n_r$ and the opposite is also true. But if this is the case, then one has $\Delta = 0$ for all $r$, and there is no anisotropy. Accordingly we should have  $n_\perp \neq n_r$  and  $K_\perp \neq K_r$.

It is worth noticing that it is not possible to have an overall anisotropy of the form $\Delta = K\rho^\gamma$, since condition $\Delta(0) = 0$  would imply  $\rho_c = 0$, which is clearly unadmissible from elementary physics considerations.

It is a simple matter to check that regularity  condition (\ref{sec2_14}) is fullfilled.

\subsection{Case 2: Radial polytrope with  $n_r = \pm\infty$ and tangential polytrope with $n_\perp \neq -1,\pm\infty$}
In this case the polytropes take the form
\begin{eqnarray}
    P_r\, &=& \, K_r\rho^{\gamma_r} \;=\; P_{rc}\, e^{-\omega}\,, \label{sec03_06}\\
    P_\perp\, &=& \, K_\perp \rho^{\gamma_\perp} \;=\; P_{\perp c}\, \omega^{1+n_\perp} \,, \label{sec03_07} 
\end{eqnarray}
where we have used the change of variables $\rho = \rho_c e^{-\omega}$ in (\ref{sec2_10}) for $P_r$, and $\rho = \rho_c\omega^{n_\perp}$ in (\ref{sec2_06}) for $P_\perp$. 

From Eqs. (\ref{sec03_06}) and (\ref{sec03_07}) we have $P_{rc} = K_r\rho_c$ and $P_{\perp c} = K_\perp \rho_c^{1+1/n_\perp}$.

Using (\ref{sec03_06}) and (\ref{sec03_07}) we obtain for  the anisotropy factor
\begin{equation}\label{sec03_07a}
    \Delta = P_\perp - P_r = P_{\perp c}\, \omega^{1+n_\perp} -
    P_{rc}\, e^{-\omega}.
\end{equation}
Then the  condition at the center $\Delta(0) = 0$ produces
\begin{equation}\label{sec03_08}
    P_{\perp c} \;=\; P_{rc} \hspace{.5cm} \longleftrightarrow \hspace{.5cm} \frac{K_\perp}{K_r} 
        \;=\; \frac{1}{\rho_c^{1/n_\perp}}\;,\;\;
\end{equation}
where we have used the condition $\omega(0) = 0$ for  the radial pressure and $\omega(0) = 1$ for the tangential one. It is a simple matter to check that  the condition  $\omega'(0)=0$ is satisfied.

\subsection{Case 3: Radial polytrope with $n_r\neq -1,\pm\infty$ and tangential polytrope with $n_\perp = \pm\infty$}
In this last case, the polytropes take the following form
\begin{eqnarray}
    P_r\, &=& \,K_r \rho^{\gamma_r} \;=\; P_{r c}\, \omega^{1+n_r} \,, \label{sec03_09}\\
    P_\perp\, &=& \, K_\perp \rho^{\gamma_\perp} \;=\; P_{\perp c}\, e^{-\omega} \,, \label{sec03_10} 
\end{eqnarray}
where we have used the change of variables $\rho =\rho_c\omega^{n_r}$ in (\ref{sec2_06}) for $P_r$, and $\rho = \rho_c e^{-\omega}$ in (\ref{sec2_10}) for $P_\perp$,  and  $P_{rc} =K_r\rho_c^{1+1/n_r}$, $P_{\perp c} = K_\perp\rho_c$.

Using the above equations  the anisotropy factor for this case reads
\begin{equation}\label{sec03_10a}
    \Delta = P_\perp - P_r = P_{\perp c}\, e^{-\omega} -
    P_{r c}\, \omega^{1+n_r}\,,
\end{equation}
and the condition at the center $\Delta(0) = 0$ produces
\begin{equation}\label{sec03_11}
    P_{\perp c} \;=\; P_{rc} \hspace{.5cm} \longleftrightarrow \hspace{.5cm} \frac{K_\perp}{K_r} 
        \;=\; \rho_c^{1/n_r}\;,\;\;
\end{equation}
where we have used the conditions $\omega(0) = 1$ for  the radial pressure term and $\omega(0) = 0$ for  the tangential one. Also it is a simple matter to check   that $\omega'(0) = 0$.

It is  worth  emphasizing that the case $n_{r,\perp} = \pm\infty$, corresponds to the  isotropic pressure case. This can  be seen from equations (\ref{sec03_07}), (\ref{sec03_11}) which imply $K_\perp = K_r$. 

\section{Solving the Anisotropic Lane--Emden Equation}\label{sec:4}
We are now ready to write down the Lane--Emden equation for the three cases considered in the previous section and to solve them.

\noindent \textbf{Case 1:} If $n _{r,\perp}\neq -1,\, \pm\infty$, the anisotropy factor and its derivative are
\begin{eqnarray}
    \Delta &=& P_{r 0} [\omega^{1+n_\perp} - \omega^{1+n_r}]\;, \label{sec4_01}\\
    \Delta' &=& P_{r 0} [(1+n_\perp)\omega^{n_\perp} -
        (1+n_r)\omega^{n_r}]\omega'\;. \label{sec4_02}
\end{eqnarray}
Then  using the Eqs. (\ref{sec4_01}) and (\ref{sec4_02}) and noticing that  the polytropic index $n$ in Eq. (\ref{sec2_08}) is in fact $n_r$ we get
\begin{eqnarray}
    \omega'' + \Pi_1\, \omega' + \Pi_2\, \omega^{n_r} \;=\; 0 \label{sec4_03}
\end{eqnarray}
with 
\begin{eqnarray}
    \Pi_1 &=& \frac{2}{z} +  \frac{2}{(1+n_r)z} 
        - \frac{2(1+\delta)}{(1+n_r)z}\, \omega^\delta\;, \label{sec4_04}\\
    \Pi_2 &=& \varepsilon + \frac{2}{(1+n_r)z^2}\, \omega^{1-n_r} 
        - \frac{2}{(1+n_r)z^2}\, \omega^{1-n_r+\delta}\;.\;\;\;\; \label{sec4_05}
\end{eqnarray}
The parameter $\delta = n_\perp - n_r$ measures  the degree of anisotropy in the system, being $\delta = 0$ the isotropic limit. It is easy to see that at the isotropic limit we obtain $\Pi_1 = 2/z$, $\Pi_2 = \varepsilon$ and the well known Lane--Emden equation is recovered.

\begin{figure}[h]
\centering
\includegraphics[width=0.48\textwidth]{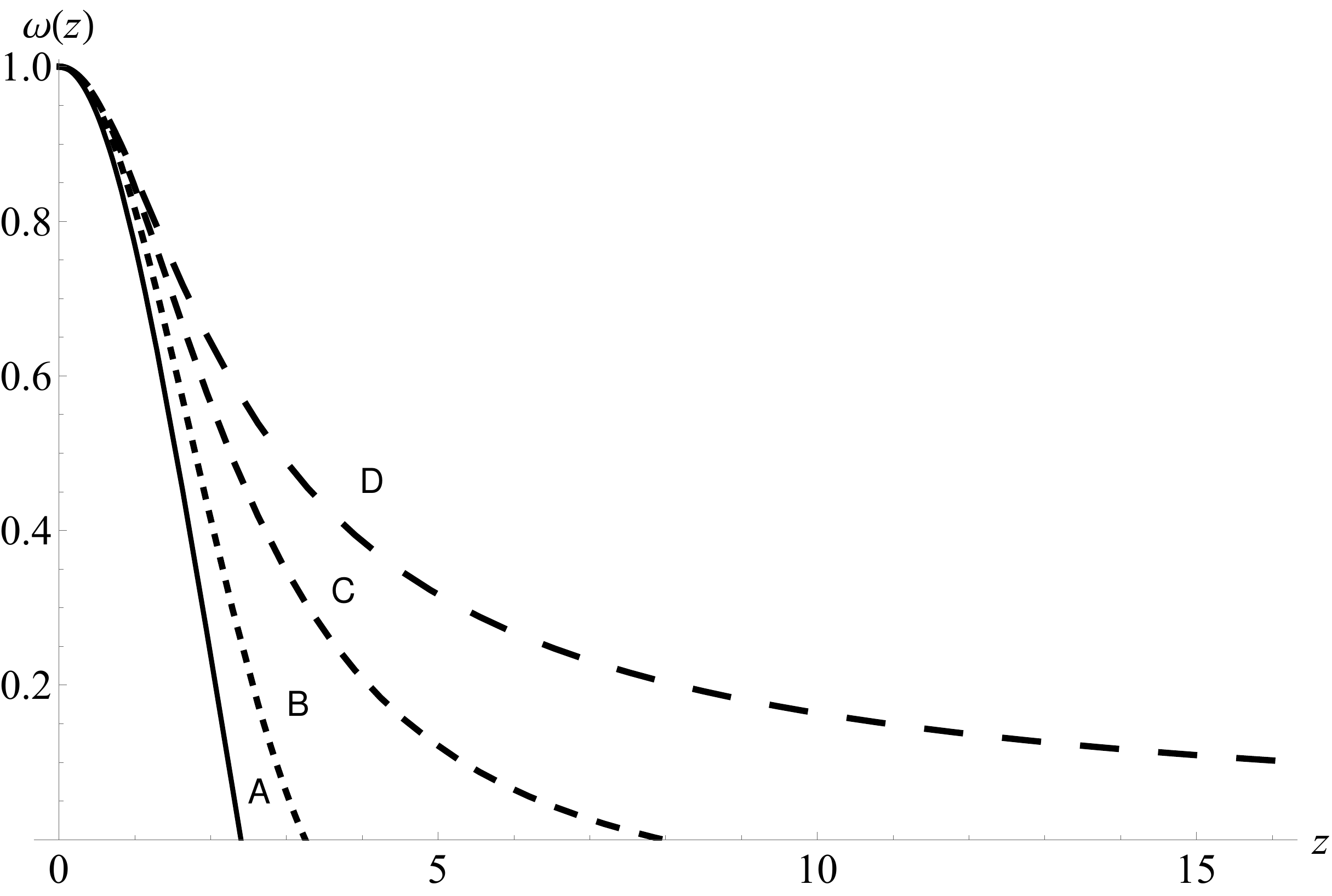} \
\caption{{\bf Case 1: } $\omega$ vs. $z$ with $\delta = 1$ and:
A) $n_r = 0$ (solid line),
B) $n_r = 1$ (small--dashed line),
C) $n_r = 3$ (medium--dashed line),
D) $n_r = 5$ (large--dashed line).
}
\label{fig:solutions01}
\end{figure}

\noindent \textbf{Case 2:} If $n _r= \pm\infty$ and  $n _\perp \neq -1,\, \pm\infty$, we obtain
\begin{eqnarray}
    \Delta &=& P_{r 0} [\omega^{1+n_\perp} - e^{-\omega}]\;, \label{sec4_06}\\
    \Delta' &=& P_{r 0} [(1+n_\perp)\omega^{n_\perp} +
        e^{-\omega}]\,\omega'\;, \label{sec4_07}
\end{eqnarray}
leading to 
\begin{eqnarray}
    \omega'' + \chi_1\, \omega' + \chi_2 \;=\; e^{-\omega}\;, \label{sec4_08}
\end{eqnarray}
where 
\begin{eqnarray}
    \chi_1 &=& \frac{2}{z} +  \frac{2}{z} 
        \left(1 + n_\perp + \omega \right)\, e^{\omega}\omega^{n_\perp} \;, \label{sec4_09}\\
    \chi_2 &=& -\frac{2}{z^2}
        \left(1 - e^{\omega} \omega^{1+n_\perp}\right) \;.\;\;\;\; \label{sec4_10}
\end{eqnarray}
It is worth noticing that, unlike the previous case, in this particular case the isotropic limit is not reachable.

\begin{figure}[h]
\centering
\includegraphics[width=0.48\textwidth]{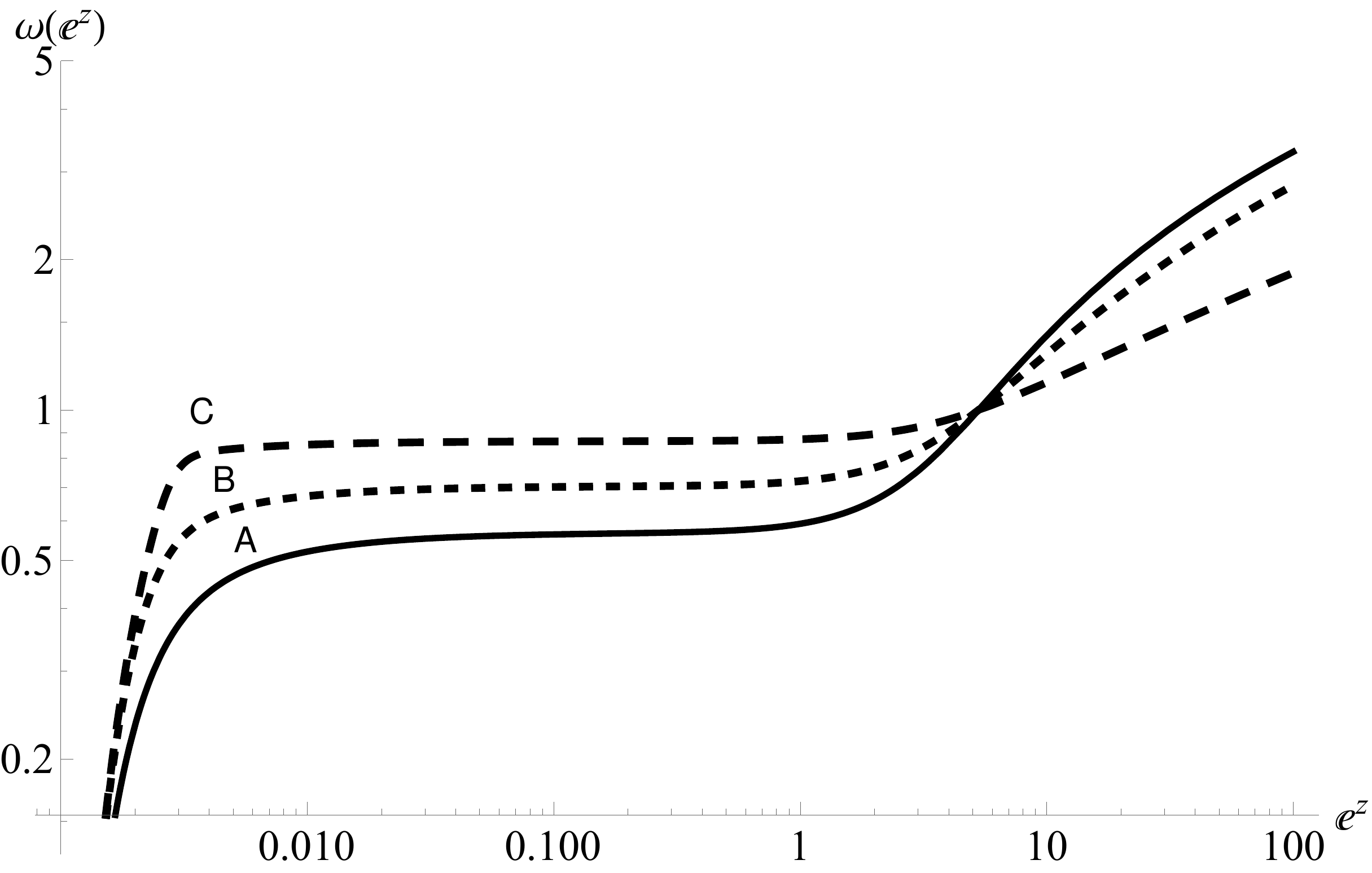} \ 
\caption{
{\bf Case 2: } $\omega$ vs. $z$ with $n_r = \pm\infty$ and:
A) $n_\perp = 0$ (solid line),
B) $n_\perp = 1$ (small--dashed line),
C) $n_\perp = 5$ (medium--dashed line),
}
\label{fig:solutions02}
\end{figure}

\noindent \textbf{Case 3:} If $n_r \neq -1,\, \pm\infty$, and  $n_\perp = \pm\infty$, the anisotropy factor and its derivative are
\begin{eqnarray}
    \Delta &=& P_{r 0} [e^{-\omega} - \omega^{1+n_r}]\;, \label{sec4_11}\\
    \Delta' &=& - P_{r 0} [e^{-\omega} + (1+n_r)\omega^{n_r}] \,\omega'\;, \label{sec4_12}
\end{eqnarray}
from which  we find
\begin{eqnarray}
    \omega'' + \Xi_1\, \omega' + \Xi_2\, \omega^{n_r} \;=\; 0\;, \label{sec4_13}
\end{eqnarray}
where the coefficients are
\begin{eqnarray}
    \Xi_1 &=& \frac{2}{z} +  \frac{2}{(1+n_r) z} + 
        \frac{2}{(1+n_r) z} \frac{e^{-\omega}}{\omega^{n_r}} +
        \frac{2n_r}{(1+n_r) z} \frac{e^{-\omega}}{\omega^{1+n_r}} \,, \nonumber\\
    \label{sec4_14}\\
    \Xi_2 &=& \varepsilon +
        \frac{2}{(1+n_r) z^2} \omega^{1-n_r} -
        \frac{2}{(1+n_r) z^2} \frac{e^{-\omega}}{\omega^{2n_r}}
    \;.\;\;\;\; \label{sec4_15}
\end{eqnarray}
As in the case 2 above, the isotropic limit is not reachable in this case.

We may now proceed to integrate numerically the obtained  equations in each case using the appropriate boundary conditions.

\begin{figure}[h]
\centering
\includegraphics[width=0.48\textwidth]{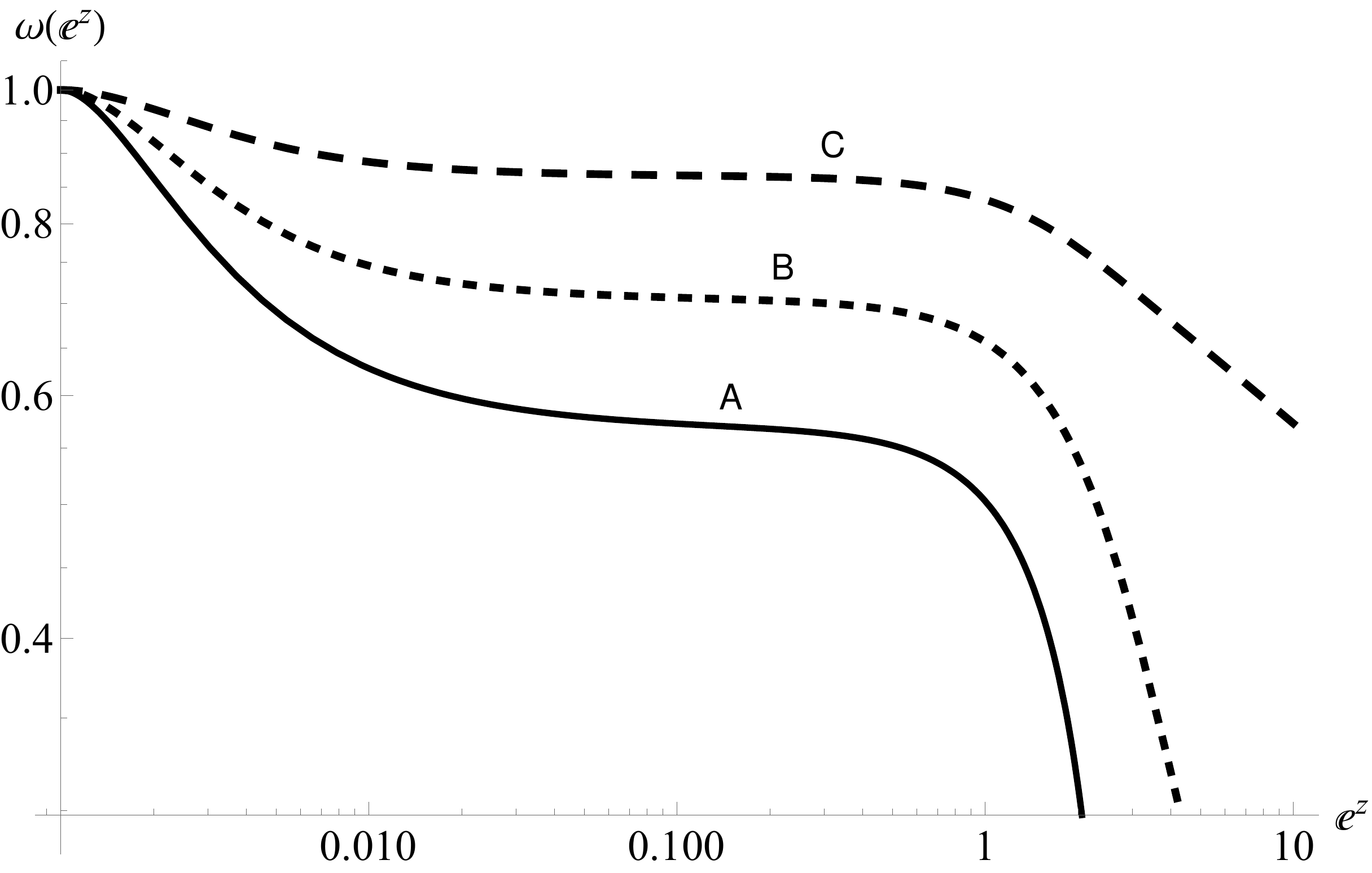} \
\caption{
{\bf Case 3: } $\omega$ vs. $z$ with $n_\perp = \pm\infty$ and:
A) $n_r = 0$ (solid line),
B) $n_r = 1$ (small--dashed line),
C) $n_r = 5$ (medium--dashed line).
}
\label{fig:solutions03}
\end{figure}

\begin{figure}[h]
\centering
\includegraphics[width=0.48\textwidth]{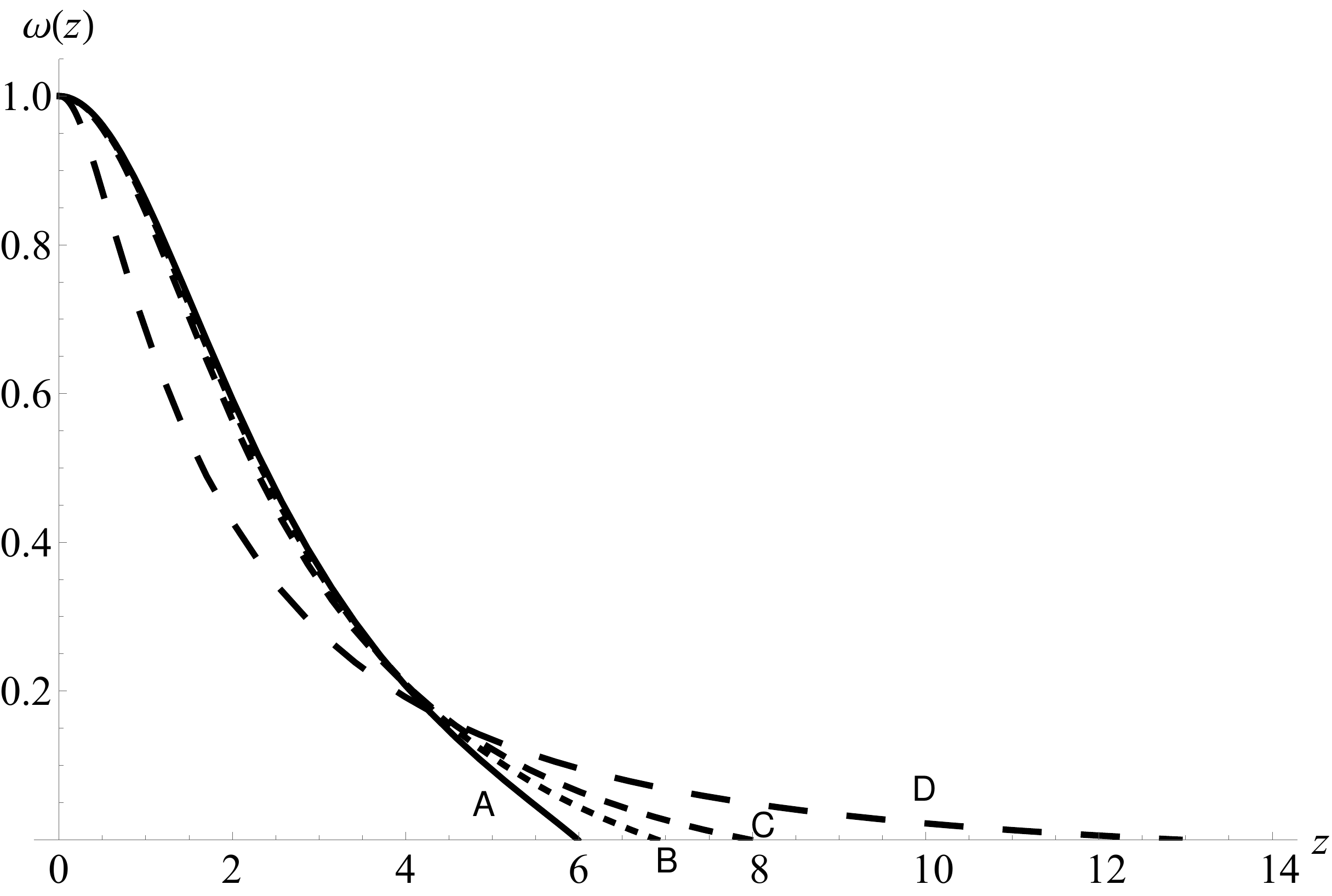}   
\caption{$\omega$ vs. $z$ with $n_r=3$ and:
A) $\delta = -0.5$ (solid line),
B) $\delta = 0$ (small--dashed line),
C) $\delta = 1$ (medium--dashed line),
iv) $\delta = 10$ (large--dashed line),
}
\label{fig:varAni01}
\end{figure}

\begin{figure}[h]
\centering
\includegraphics[width=0.48\textwidth]{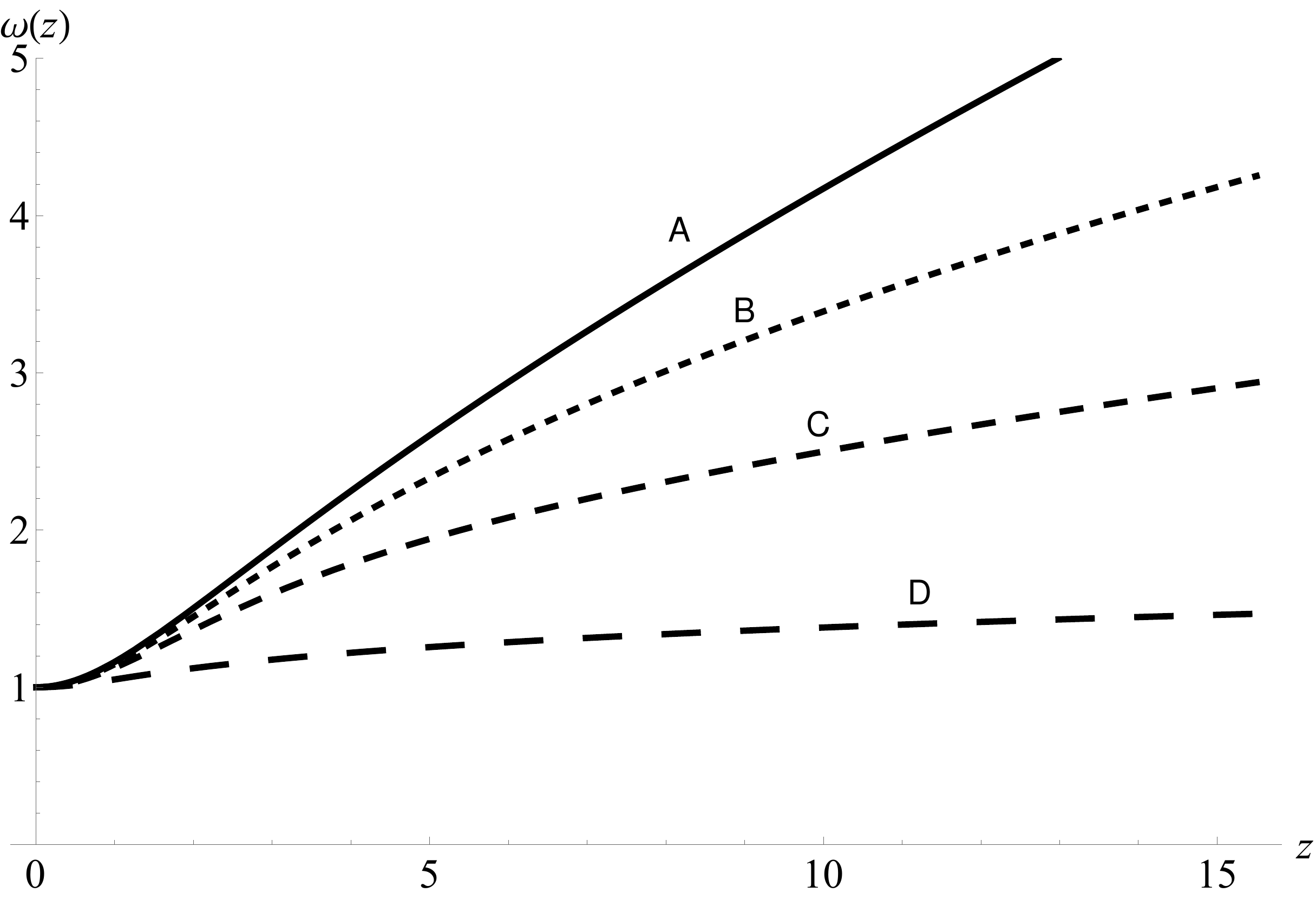}   
\caption{$\omega$ vs. $z$ with $n_r=-3$ and:
A) $\delta = -0.5$ (solid line),
B) $\delta = 0$ (small--dashed line),
C) $\delta = 1$ (medium--dashed line),
D) $\delta = 10$ (large--dashed line),
}
\label{fig:varAni02}
\end{figure}

Figures \ref{fig:solutions01}--\ref{fig:solutions03} show the solutions for the three different cases and  a selected range of values of the parameters, whereas Figures \ref{fig:varAni01}--\ref{fig:varAni02} exhibit the behavior of some models of the case 1 for the indicated range of values of the parameters.

One of the most interesting problem we have to deal with,  within the context of the formalism developed here, concerns the modifications introduced by the anisotropy of the pressure in the Chandrasekhar mass. Since we would like to compare the results with the isotropic case we shall restrict to the case 1. 

The total mass of the distribution is given by the following expression
\begin{equation}\label{sec4_16}
    M \;=\; 4\pi \int_0^R \! r^2\rho \,dr \;=\; 
    4\pi \rho_c \frac{R^3}{(z_n^{(\delta)})^3} \int_0^{z_n^{(\delta)}} \! z^2\omega^{n_r} \,dz   \;,
\end{equation}
were we have  used  (\ref{sec2_06}), and $z_n^{(\delta)}$ satisfies the relation $\omega(z_n^{(\delta)}) = \omega_n^{(\delta)} = 0$, where $\omega(z_n^{(\delta)})$ is the solution of (\ref{sec4_03}) for a given value of $\delta$ and $n_r$. Using the Eq. (\ref{sec2_08}) we find  for the integrand, 
\begin{eqnarray}\label{sec4_17}
    \varepsilon z^2\omega^{n_r } &=& 
    \frac{2z}{P_c(1+n_r)\omega^{n_r}}\!
    \left(\! \Delta' + \frac{\Delta}{z} - n_r \frac{\omega'}{\omega}\Delta \! \right)\nonumber\\
    & & - \frac{d}{dz}\left( z^2 \frac{d\omega}{dz} \right) \!.
\end{eqnarray}
Substituting this expression in (\ref{sec4_16}) and using (\ref{sec4_01}) and (\ref{sec4_02}) we find that the total mass is
\begin{eqnarray}\label{sec4_18}
        \!\!\! M &=& 4\pi \rho_c \frac{R^3}{\varepsilon (z_n^{(\delta)})^3} \bigg[ 
        -\!\! \int_0^{z_n^{(\delta)}} \!\! \frac{d}{dz}\left( z^2 \frac{d\omega}{dz} \right) dz 
        \nonumber \\
        & & +\; \frac{2}{P_c(1+n_r)}
        \int_0^{z_n^{(\delta)}} \!\!\! \frac{z}{\omega^{n_r}}\! \left(\! \Delta' + \frac{\Delta}{z} - n_r \frac{\omega'}{\omega}\Delta \! \right)dz\bigg] \nonumber \\
        &=& 4\pi \rho_c \frac{R^3}{\varepsilon (z_n^{(\delta)})^3} \bigg\{ 
        -(z^2\omega')_{z_n^{(\delta)}} \nonumber \\
        & & + \; \frac{2}{(1+n_r)} \!
        \int_0^{z_n^{(\delta)}} \!\!\!\! z\! \left[\! (1+\delta)\omega^{\delta}\omega' \!
        + \frac{\omega^{1+\delta}}{z} -\omega' - \frac{\omega}{z} \right]\!dz\bigg\}  \nonumber \\
        &=& 4\pi \rho_c \frac{R^3}{\varepsilon (z_n^{(\delta)})^3} \bigg[ 
        -(z^2\omega')_{z_n^{(\delta)}} \nonumber \\
        & & +\; \frac{2}{(1+n_r)} \!
        \int_0^{z_n^{(\delta)}} \!\!\!\! \frac{d}{dz}\! \left(\! z \omega^{1+\delta} 
        - z\omega \right)\!dz\bigg] \nonumber \\
        &=& 4\pi \rho_c \frac{R^3}{\varepsilon (z_n^{(\delta)})^3} \bigg[ 
        -(z^2\omega')_{z_n^{(\delta)}} \bigg] .     
\end{eqnarray}
In the last line we have used the condition $\omega(z_n^{(\delta)}) = 0$. Note that Chandrasekhar mass $M_{ch}$ is obtained from (\ref{sec4_18}) using parameters $\delta = 0$ and $n = n_r = 3$. In order to compare the anisotropic mass $M$ with $M_{ch}$ we focus in models that are connected with Chandrasekhar model, so setting $\varepsilon = +1$ we find that
\begin{equation}\label{sec4_19}
    \frac{M}{M_{ch}} \;=\; 
    \frac{(-z^2\omega')_{z_n^{(\delta)}}}{(-z^2\omega')_{z_3^{(0)}}}\;.
\end{equation}
If we introduce the mean density $\Bar{\rho} = 3M/4\pi R^3$, we can study the density concentration
\begin{equation}\label{sec4_20}
    \frac{\bar{\rho}}{\rho_c} \;=\; \left( -\frac{3}{z} \omega' \right)_{z_n^{(\delta)}}\,.
\end{equation}
This is a useful expression because it depends only of the model parameters.

\begin{figure}[h]
\centering
\includegraphics[width=0.48\textwidth]{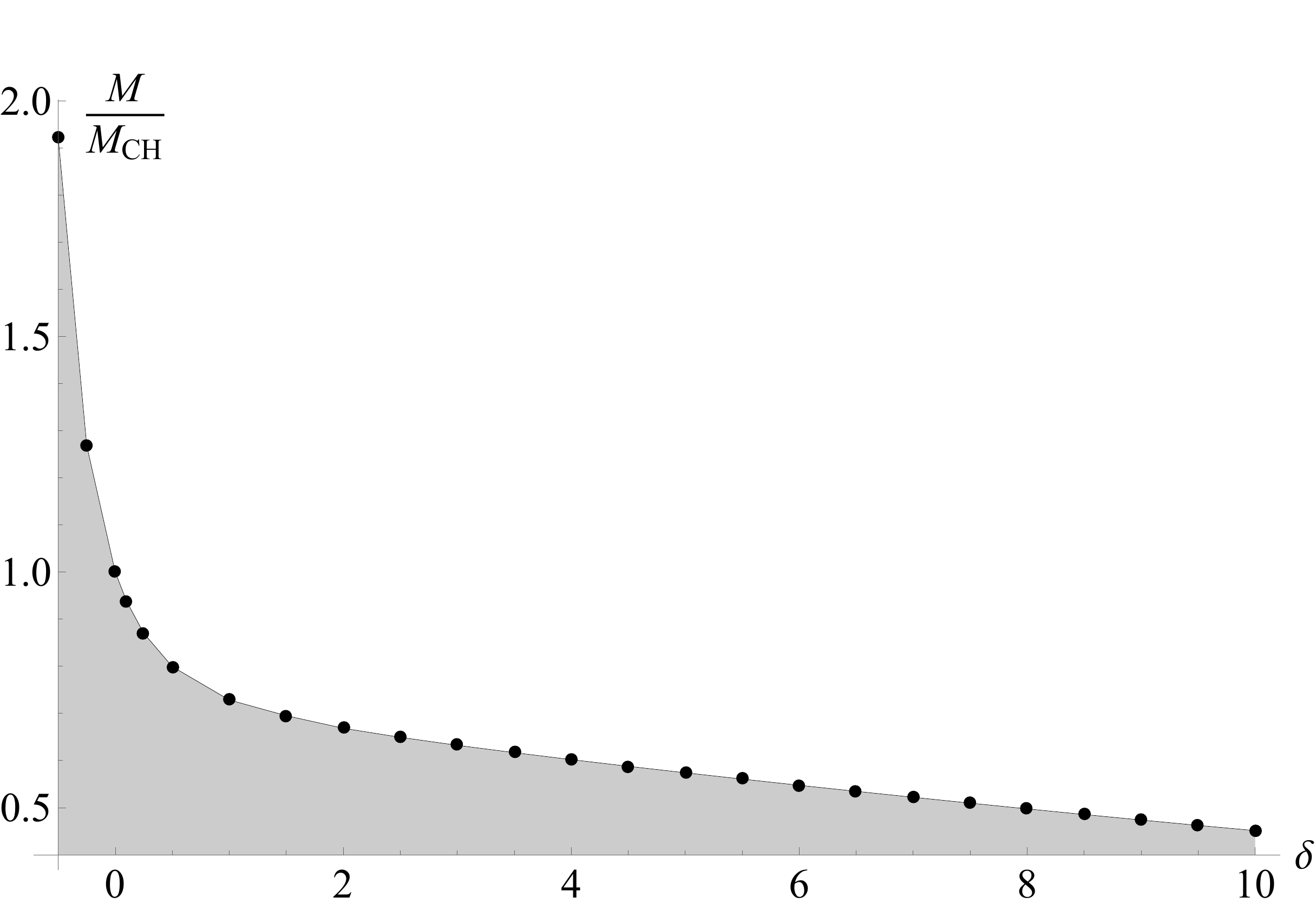}   
\caption{Mass ratio $M/M_{ch}$ vs. anisotropy parameter $\delta$.
}
\label{fig:mass-scaling}
\end{figure}

In table \ref{tab:values} we give a set of numbers that can be used to create concrete models. Note that for a fixed radial polytropic index $n_r = 3$, when the anisotropy parameter $\delta$ increases, the density concentration becomes smaller. Using this numerical values, we obtain a plot of Eq. (\ref{sec4_19}) versus the anisotropy parameter $\delta$, Fig. \ref{fig:mass-scaling}. From this figure it is easy to see how the mass can be stretched or shrunk depending on the value of the anisotropy parameter.

\begin{table}[h]
\centering
 \begin{tabular}{c c c c} 
 \hline\hline
 $\delta$ & $z_3^{(\delta)}$ & $(-z^2\omega')_{z_3^{(\delta)}}$ & $\rho_c/\bar{\rho}$ \\
 \hline\hline
 0 & 6.89684 & 2.01824 & 54.1824 \\ 
 0.5 & 7.55491 & 1.61324 & 89.0983 \\
 1.5 & 8.17633 & 1.40294 & 129.871 \\
 2.5 & 8.61012 & 1.31038 & 162.37 \\
 3.5 & 8.93913 & 1.24468 & 191.296 \\
 4.5 & 9.29552 & 1.18622 & 225.701 \\
 7.5 & 10.806 & 1.02843 & 408.983 \\
 10.5 & 13.4609 & 0.888946 & 914.595 \\
 20.5 & 36.1659 & 0.634237 & 24861.3 \\
 50.0 & 172.18 & 0.53336 & $3.19203\cdot 10^6$ \\
 100.0 & 406.918 & 0.516845 & $4.3485\cdot 10^7$ \\
 \hline
 \end{tabular}\label{tab:values}
 \caption{Numerical values for polytropic models with $\varepsilon = 1$ and $n_r = 3$.}
\end{table}

\section{Concluding Remarks}\label{final}
We have described a whole family of polytropes for anisotropic matter by assuming that both pressures satisfy a  polytropic equation of state.  The main advantage of the approach followed here with respect to the kind of anisotropy considered in \cite{Herrera:2013dfa} resides in the fact that in our case we know that, at least,  for some range of values of $\delta$ not very far from the isotropy, our equation of state (\ref{sec03_02}) is well justified. 

We have established the structure equations (Lane--Emden) describing each  model for any set of the parameters. Next, we have applied a graphical-numerical treatment to solve the generalized Lane-Emden equations that arise in each case and have calculated the Chandrasekhar mass for a white dwarf. It is clearly shown  that the Chandrasekhar mass limit changes with the introduced anisotropy. These models can be further developed and used to study the influence of local anisotropy in such an important problem as the Chandrasekhar mass limit, in particular in relation with the possible existence of super-Chandrasekhar white dwarfs.  At this point we cannot assert   if the inferred super--Chandrasekhar white dwarfs from collected data \cite{howell2006,scalzo2010,scalzo2012,hachisu2012,das2013} are the result of anisotropy as considered here, but this interesting issue deserves more attention.  

The case  $\gamma=1$ deserves special attention. Indeed, we know that in the isotropic pressure case, polytropes with  $\gamma=1$ are used to construct models with non--degenerate isothermal cores, which play an important role in the analysis of  the Schonberg-Chandrasekhar limit.  Let us recall that after the hydrogen burning in a main sequence  star, the stability of the resulting helium core surrounded by a hydrogen--rich envelope is of the utmost relevance to  predict the subsequent evolution of the star (its place in the Hertzprung--Russell diagram).  The Schonberg-Chandrasekhar limit concerns the ratio $M/M_c$ where M is the total mass and $M_c$ is the mass of the core, and asserts that such a ratio must not exceed a certain limiting value (the Schonberg-Chandrasekhar limit), for otherwise the system is no longer stable. Now, the important point is that to arrive at this result one has to resort to the virial theorem, which we know is affected by the presence of pressure anisotropy (see pages 95, 96  in \cite{Herrera:1997plx} for a discussion on this point). In other words pressure anisotropy would affect the Schonberg-Chandrasekhar limit in two different ways. On the one hand by affecting the structure of the polytrope and on the other by the modifications of the virial theorem, introduced by the pressure anisotropy.  To establish how this limit is specifically affected by the kind  of the  anisotropy considered here, is out of the scope of this work, but certainly is an issue that should be addressed in the future.

Finally, we have to point out that this study was carried out  within the context of Newtonian gravity and spherical symmetry. It is possible that this symmetry could be broken by the same physical factors that create the anisotropy in the system, in which case the method presented here should be applied with caution and only as an approximation.


\bibliographystyle{ieeetr}
\bibliography{biblioanipolynew.bib}

\end{document}